\documentclass{aastex}
\usepackage{spr-astr-addons}
\usepackage{url}
\usepackage{graphics}

\begin{document}

\title{Linear line spectropolarimetry of Herbig Ae/Be stars}
\shorttitle{Linear line Spectropolarimetry of Herbig Ae/Be stars}
\shortauthors{Jorick S. Vink}

\author{Jorick S. Vink\altaffilmark{1}} 
\affil{Armagh Observatory, College Hill, BT61 9DG Armagh, Northern Ireland}

\begin{abstract}
Accretion is the prime mode of star formation, but 
the exact mode has not yet been identified in 
the Herbig Ae/Be mass range.
We provide evidence that the the maximum variation 
in mass-accretion rate is reached on a 
rotational timescale, which suggests that rotational modulation 
is the key to understanding mass accretion. 
We show how spectropolarimetry is uniquely capable of resolving 
the innermost (within 0.1 AU) regions between the star and the disk, 
allowing us to map the 3D geometry of the accreting gas, and test 
theories of angular momentum evolution. 
We present Monte Carlo line-emission simulations showing how 
one would observe changes in the polarization properties on 
rotational timescales, as accretion columns come and go into 
our line of sight. 
\end{abstract}



\section{Introduction}

Herbig Ae/Be stars with masses in the range 2-15\,$M_{\odot}$ 
lie at the interface between low-mass and high-mass 
star formation. One of the key goals is to unravel whether the magnetospheric accretion
model that is very successfully applied in low-mass ($M < 2 M_{\odot}$) 
T Tauri stars may also be of relevance for  
higher mass stars. Whilst {\it circular} Stokes $V$ spectropolarimetry can be used to measure magnetic fields, 
{\it linear} Stokes $QU$ polarimetry may be employed to probe the gas geometry within  
the innermost regions between the star and the disk -- on the scale of just a few stellar radii. 

Such work is needed to unravel the complex accretion flows onto T Tauri and Herbig stars, in order to understand 
issues such as the mass-accretion rate $\dot{M}_{\rm acc}$ versus stellar mass 
relation (Garcia-Lopez et al. 2006; Mendigutia et al. 2012) as well as the more general questions 
as to whether even the highest mass stars in the Universe form by disk accretion or 
whether more exotic formation mechanisms, involving for instance stellar collisions, need to 
be invoked (e.g. Bestenlehner et al. 2011). 

\begin{figure}
\centerline{\includegraphics[width=8cm]{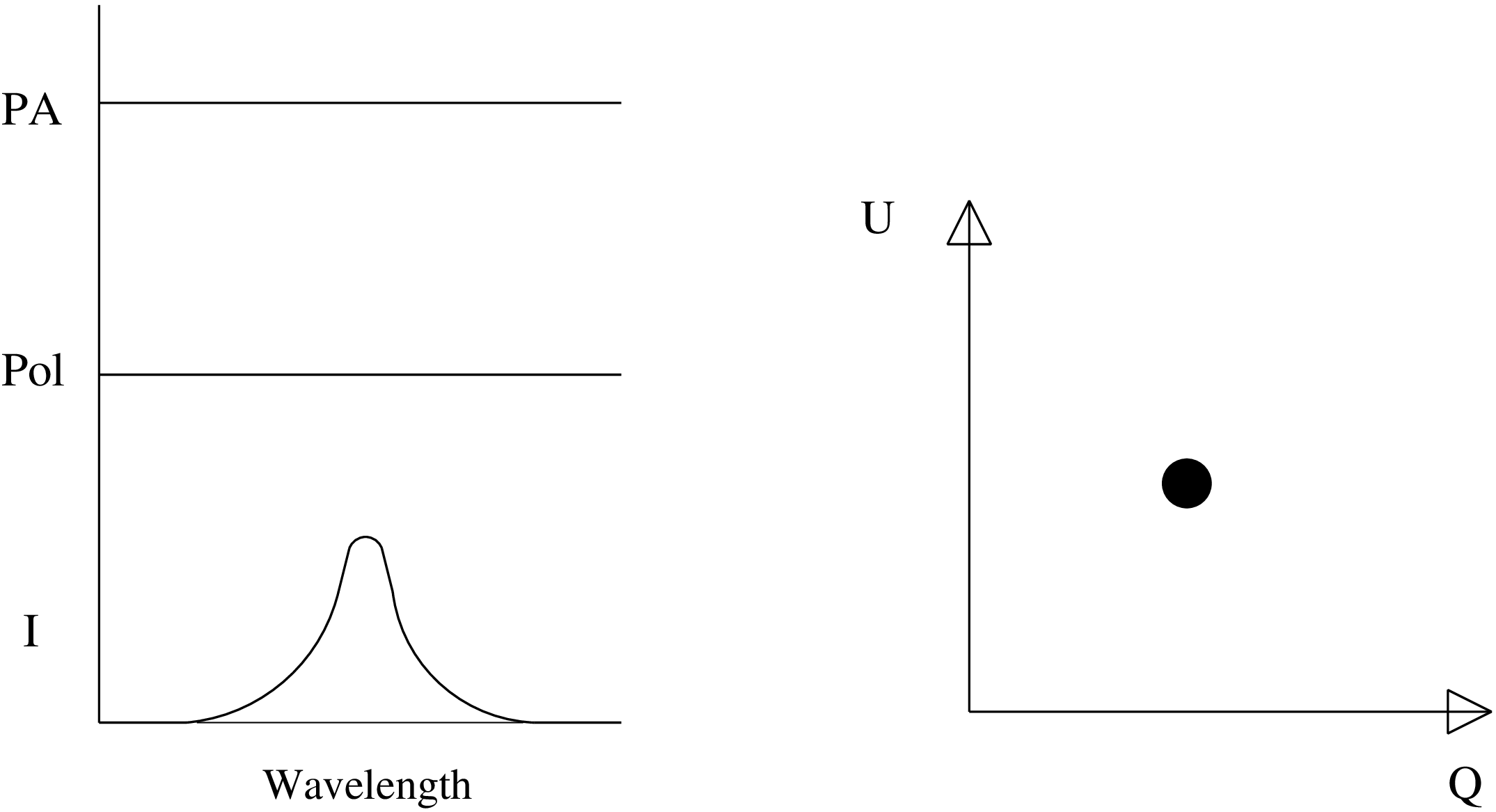}}
\caption{Cartoon indicating the simplest case of {\it no line effect}. On the left, 
the polarization spectrum {\it triplot}  -- and on the right its accompanying $QU$ diagram.
A typical emission is shown in the lower panel of the triplot, 
the \%Pol is given in the middle panel, while the Position Angle (PA) is 
shown in the upper panel of the triplot. See Vink et al. (2002) for further details.}
\label{noline}
\end{figure}

We will explore the alignment between disk position angles (PAs) 
measured from linear $QU$ spectropolarimetry and those from Herbig binaries, showing that our data are
fully consistent with disk accretion and fragmentation in the intermediate mass 2-15\,$M_{\odot}$ (Harmanec 1988) range. Note that
 this is independent of the polarizing mechanism in Herbig Ae/Be stars.  

\section{The method}

Linear Stokes $QU$ polarimetry can 
be used to measure flattening of the circumstellar medium. 
In principle, continuum polarization would already be able to inform us about the presence 
of an asymmetric disk-like structure on the sky, but in practice this 
issue is complicated by the roles of intervening circumstellar (as e.g. in the UXOR phenomenon; Grinin 1994) and/or interstellar 
dust, as well as instrumental polarization. This is one of the reasons why {\it spectro}polarimetry, measuring
the change in the degree of linear polarization across spectral lines is such a powerful tool, as {\it intrinsic} 
information can directly be obtained from the Stokes $QU$ plane. 
The second reason is the additional bonus that it may provide kinematic information
of the flows around young stars.

\begin{figure}
\centerline{\includegraphics[width=8cm]{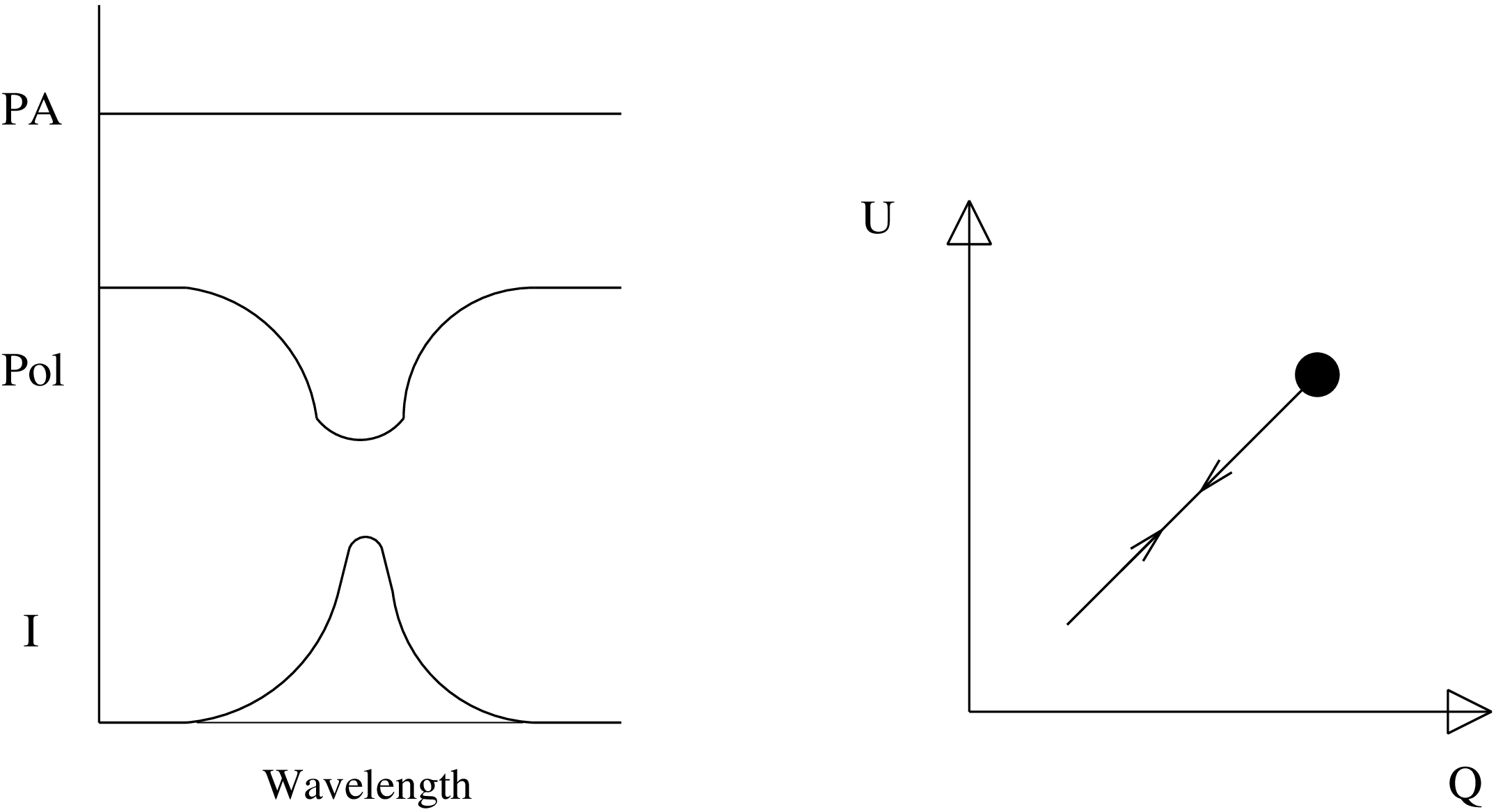}}
\caption{Cartoon indication {\it depolarization} or {\it dilution} of the emission line. 
The depolarization across the line is as broad as the Stokes $I$ emission. Depolarization
translates into Stokes $QU$ space as a linear excursion. See Vink et al. (2002).}
\label{depolariz}
\end{figure}

Figures 1-3 show spectropolarimetry
cartoons (both in terms of polarization ``triplot'' spectra and $QU$ planes) 
for the case that a spatially unresolved object is (i) 
spherically symmetric on the sky with ``no line effect'', (ii) asymmetric  
subject to line ``depolarization'' where the emission line simply acts to ``dilute'' the polarized 
continuum, or (iii) where the line effect is more subtle, involving 
PA flips across intrinsically polarized emission lines.

In its simplest form, spectropolarimetry can be used to detect a difference 
between an unpolarized emission line such as H$\alpha$ and a polarized continuum that results 
from scattering off a circumstellar disk (Fig.~\ref{depolariz}). This tool has for instance 
been employed on samples of classical Be stars in the 1970s (e.g. Poeckert \& Marlborough 1976). These 
observations gave an incidence rate of order 60\%, which is fully consistent with {\it all} classical Be stars
being embedded in electron scattering disks. Given that the PA measured from linear spectropolarimetry 
was consistent with those from interferometry, the technique was considered to be a particularly efficient 
and accurate method for discovering disks around stars that would otherwise remain unresolved. 

\begin{figure}
\centerline{\includegraphics[width=8cm]{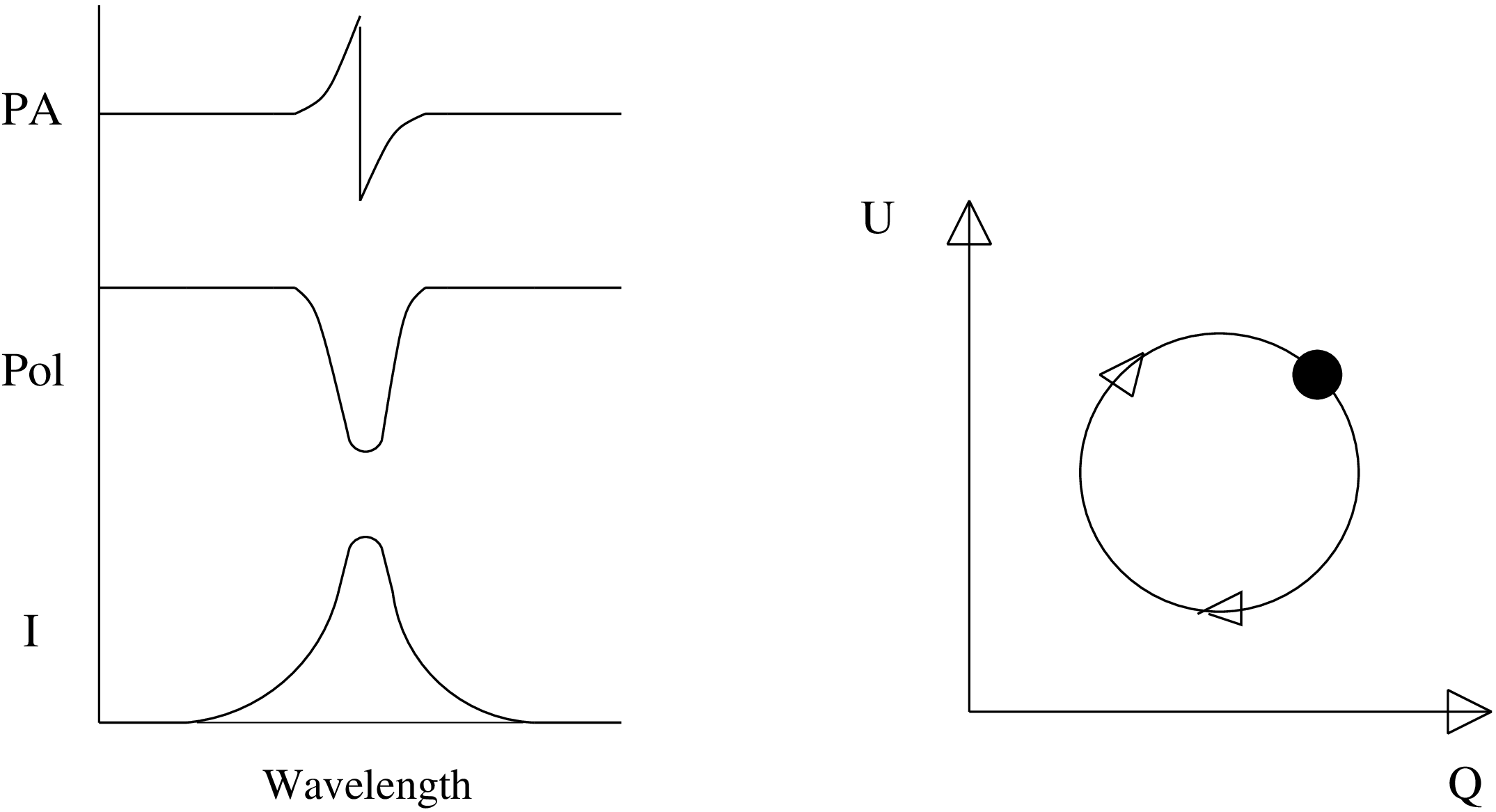}}
\caption{Cartoon showing a compact source of line photons scattered off a {\it rotating} disk. 
The polarization signatures are relatively narrow compared 
to the Stokes $I$ emission line. The PA flip is associated with a loop
in $QU$ space. See Vink et al. (2002).}
\label{narrow}
\end{figure}

In more recent years, we have found incidences of intrinsic {\it line} polarization, implying 
that it is the line itself that is polarized, e.g. by a rotating accretion disk around a T Tauri star
that scatters line photons from the interior regions close to the star (Fig.~\ref{narrow}), that may arise 
from the magnetospheric accretion model (Vink et al. 2003). 

\section{Results}

\begin{figure}
\centerline{\includegraphics[width=6cm]{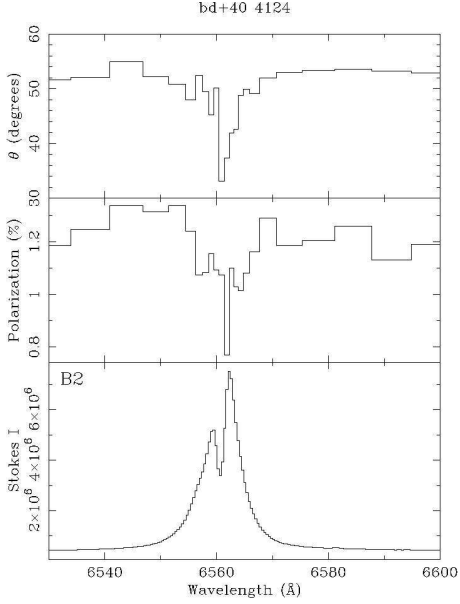}}
\caption{
The Herbig Be star BD$+$40 4124 showing line ``depolarization'' or ``dilution''. 
The {\it triplot} shows Stokes $I$ emission in the lowest panel, 
the \%Pol is given in the middle panel, while the Position Angle (PA) is 
given in the upper panel.  
The data are re-binned such that the 1$\sigma$ error in the polarization
corresponds to 0.05\% as calculated from photon statistics. 
See Vink et al. (2002) for further details.}
\label{bd40}
\end{figure}

\begin{figure}
\centerline{\includegraphics[width=6cm]{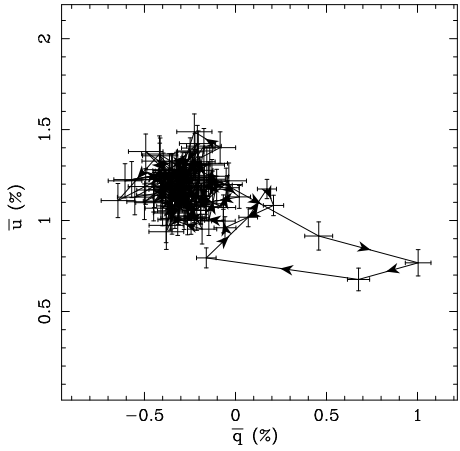}}
\caption{
$QU$ diagram of the Herbig Be star BD$+$40 4124 showing depolarization/dilution. 
See Vink et al. (2002) for further details.}
\label{qubd40}
\end{figure}

Over the last decade, we have been active in obtaining medium resolution ($R$ $\sim$ 8000) 
linear spectropolarimetry on samples of several tens of T Tauri and Herbig (single and binary) stars 
on 4m class telescopes at high signal-to-noise ratios $>$ 1000 using 
instruments such as the ISIS spectrograph on the William Herschel Telescope (WHT).  

\subsection{A transition in mode of mass accretion between Herbig Ae and Be stars}

\begin{figure}
\centerline{\includegraphics[width=6cm]{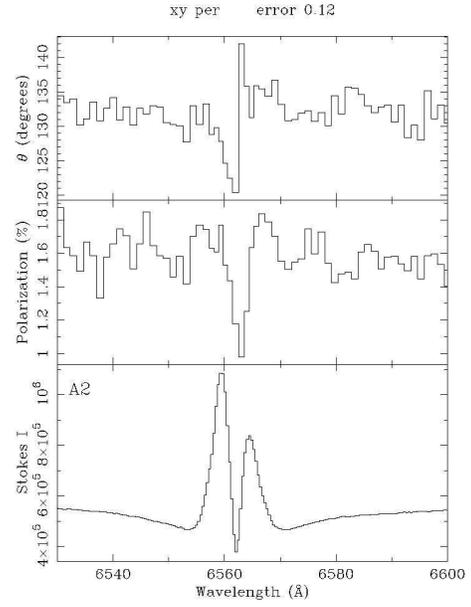}}
\caption{
The Herbig Ae star XY Per shows intrinsic {\it line} polarization. 
Note the flip in the PA, which would translate into a {\it loop} when the 
data were plotted in a $QU$ diagram. See Vink et al. (2002) for further details.
}
\label{xyper}
\end{figure}

\begin{figure}
\centerline{\includegraphics[width=6cm]{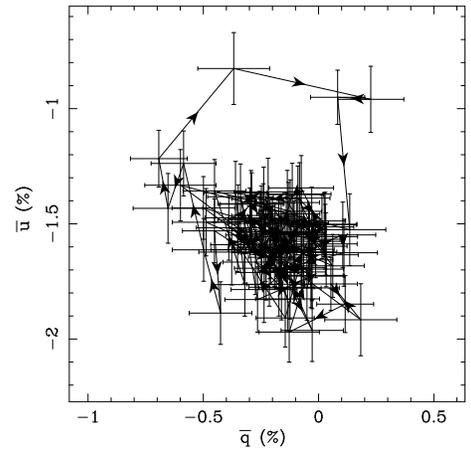}}
\caption{
$QU$ diagram of the Herbig Ae star XY Per showing a loop representative of 
intrinsic {\it line} polarization. 
See Vink et al. (2002) for further details.
}
\label{quxyper}
\end{figure}

Figures\,\ref{bd40} to\,\ref{quxyper} show the difference in linear H$\alpha$ spectropolarimetry 
discovered by Vink et al. (2002). Herbig Ae stars show PA flips in the upper panel of the 
triplot, which is caused by intrinsic line emission scattered off a rotating disk. These 
PA flip ($QU$ loop) data are the same in T Tauri stars, where the magnetospheric accretion model
has been successfully applied. The Herbig Be stars show spectropolarimetric behaviour that is 
notably different: here the data are consistent with disk accretion. Note that this
difference between the Ae and the Be stars is not confined to H$\alpha$, as 
H$\beta$ and H$\gamma$ show exactly the same differences between the Herbig Ae and the 
Herbig Be stars (Mottram et al. 2007).

\subsection{Constraining the disk inner radius}

\begin{figure}
\centerline{\includegraphics[width=9cm]{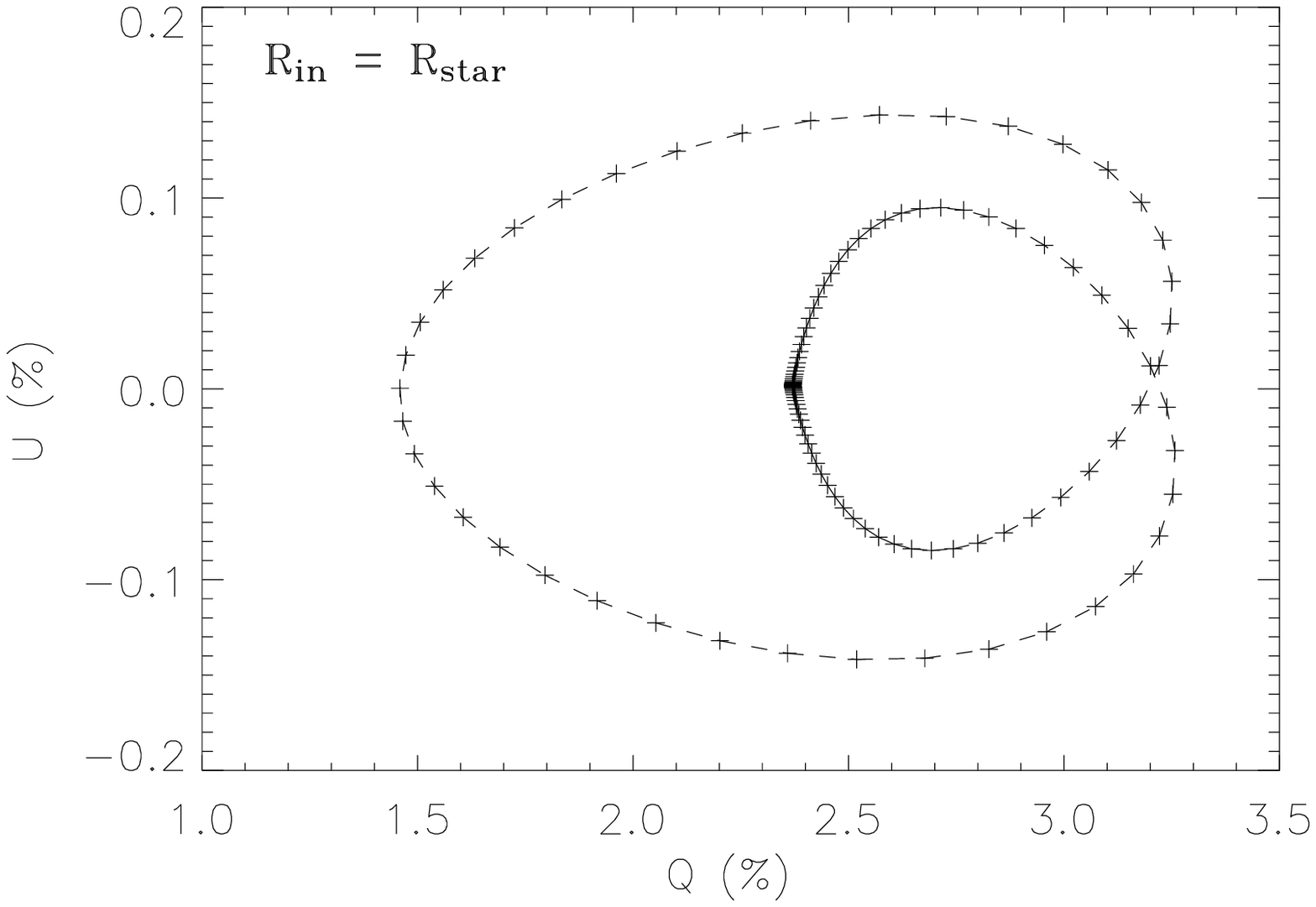}}
\caption{
Stokes $QU$ Monte Carlo model data for the case of {\it no} inner hole, subject to 
a double $QU$ loop, which could be represented by 2 PA rotations, when plotted versus wavelength, such as in Fig.\,\ref{xyper}. See Vink et al. (2005a) for more detail.
}
\label{1rstar}
\end{figure}

\begin{figure}
\centerline{\includegraphics[width=9cm]{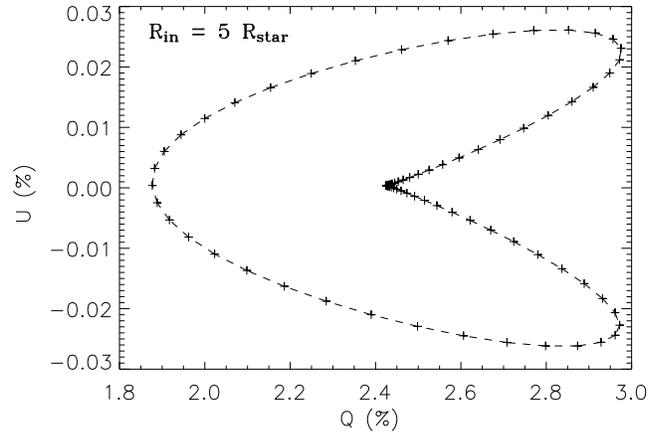}}
\caption{
Stokes $QU$ Monte Carlo model data for the case of a larger inner hole, subject to 
a single $QU$ loop, which could also be represented by a single PA rotation, when plotted versus wavelength, such as in Fig.\,\ref{xyper}. See Vink et al. (2005a) for more detail.
}
\label{5rstar}
\end{figure}

Vink et al. (2005a) performed 3D Monte Carlo scattering experiments 
using {\sc torus} (Harries 2000) 
to constrain the disk inner radius in T Tauri and Herbig Ae stars. 
The results were obtained for both a flat disk and a constant opening angle ``theta''
disk and the model objects were found to show loops in the $QU$ diagram data (corresponding to 
PA flips as shown in the upper panel of the triplot in Fig.\,\ref{xyper}). 

There is a notable difference between a disk with no inner hole 
showing {\it double QU} loops, as shown in Fig.\,\ref{1rstar}, versus 
a disk with a small inner hole, subject to {\it single} QU loops, such 
as shown in Fig.\,\ref{5rstar}.
 
\begin{figure}
\includegraphics[width=8cm]{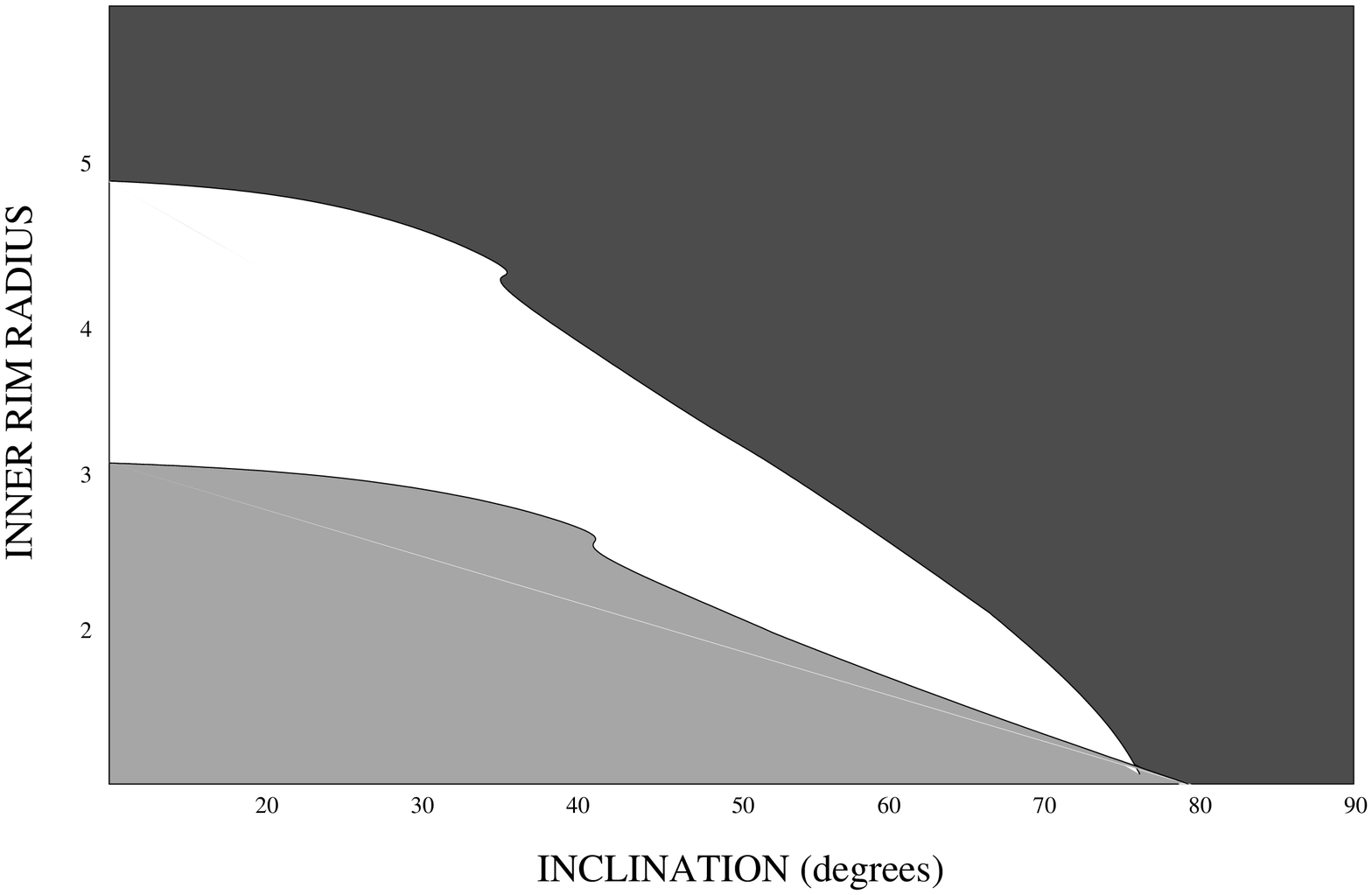}
\caption{
Constraining the disk inner hole. Single QU loops are given by the dark shaded area, whilst 
double QU loops are indicated by the light shaded areas of the disk inner radius vs. disks inclination 
plane. Transitional behaviour is represented by the white shaded area of the plane. See Vink et al. (2005).
}
\label{incl}
\end{figure}

Using this qualitatively different 
behaviour, it was shown how this methodology can be used to derive {\it quantitative} constraints on 
the sizes of disk inner holes around T Tauri and Herbig stars. 
Figure\,\ref{incl} summarizes our methodology. For instance, data on the bright 
T~Tauri star GW~Ori show the presence of a gradual PA change 
across H$\alpha$, which may be indicating the presence of a relatively small 
inner hole of less than two stellar radii for an inclined disk ($i$ $\simeq$ 75 degrees), or a 
more pole-on disk, but with a larger inner hole (Vink et al. 2005a; Vink et al. 2005b).

\subsection{Disk alignment in Herbig Ae/Be binaries}

Vink et al. (2005b) and Wheelwright et al. (2011) compared their spectropolarimetric disk 
PAs to the disk PAs derived from alternative imaging techniques, such as interferometry. 
The excellent agreement 
proved that linear spectropolarimetry is an efficient tool to determine disk PAs, and 
that this is independent of the polarizing mechanism, whether that involves 
scattering (Vink et al. 2005; Milic \& Faurobert 2014) or optical pumping (Kuhn et al. 2007).

The disk PAs were compared to binary PAs by Wheelwright et al. (2011) 
from Herbig Ae/Be binaries. Our PA data were found to be 
inconsistent with a random distribution, which might perhaps have 
been expected in case primary stars capture their companions in
a cluster-like competitive accretion scenario. Instead, the PA data were found to be
consistent with a disk accretion scenario, in which both the 
primary and secondary object fragment from the same accretion disk. 

In other words, our data suggest that stars up to 15\,$M_{\odot}$ may indeed form by disk accretion, as 
modeled by e.g. Kratter \& Matzner (2006), Krumholz et al. (2009), and Kuiper et al. (2010).

\section{Current applications and limitations}

We have shown polarimetric line profiles for scattering off rotating 
disks, but we have not yet given any preference to a particular type of 
scattering particle.
Moreover, there is usually the issue of interstellar polarization,  
implying that the observed continuum PA is generally not equal to zero and 
that the level of continuum polarization is affected by a foreground contribution
due to interstellar grains.
Nonetheless, the {\it differential} effect between line and continuum is not affected by 
foreground polarization, and the shapes of the loops in the $QU$ plane 
remain exactly the same (they are just shifted).
Furthermore, there is the complication of unpolarized line emission. 
For classical Be stars, H$\alpha$ is believed to form in the circumstellar environment, rather than at 
the stellar surface, as assumed in our Monte Carlo models. 
So, there are two potential polarization effects (dilution and intrinsic line polarization), which 
could sometimes be at work simultaneously, but the PA rotations due to intrinsic line 
polarization should be distinguishable from dilution effects because of 
their contrasting characteristics in the $QU$ plane.

So far, we have assumed idealized disk geometries. 
One may for instance also wish to consider more sophisticated flaring disk geometries, as 
infrared spectral energy distribution modeling has indicated a preference 
for this (Kenyon \& Hartmann 1987; Chiang \& Goldreich 1997. 
These flaring disks may even posses puffed-up inner rims (e.g. Dullemond et al. 2001). 
A flaring disk may be able to intercept light at larger distances from the stars, which may 
contribute to the continuum polarization percentage. However, the differences between 
a flared and constant opening angle disk are expected to be minor as far as the predicted
polarization changes across spectral line profiles are concerned.

The issue of the polarizing agent has not yet been settled. 
For hot stars, such as 
classical Be stars, the polarizing agent is usually attributed to electron scattering 
(although hydrogen continuum opacity is thought to have an effect on broad-band 
spectropolarimetry at ionization edges). Electrons are known to be 
able to smear out line polarimetric profiles because of their large thermal motions 
compared to the bulk motions of the stellar envelope (Wood \& Brown 1994).
Therefore, some of the Stokes $Q$ and $U$ structure across lines  
would be diminished for hot stars by this thermal broadening effect. This would especially be 
true for lower sensitivity measurements. To date, line polarimetry with 4m-class telescopes 
is usually performed at the S/N level of 1000, and the accuracies are therefore 
of the order of about 0.1\%. Demands 
on ``differential'' measurements across spectral lines are less severe than absolute 
ones. Already some of the published data from the WHT and the Anglo Australian Telescope (AAT) 
has performed better than at the 0.1\% level (see 
the data on Zeta Pup by Harries \& Howarth 1996).
Nonetheless, in the current era of 8m-class, and in the upcoming era of 40m-class telescopes, 
larger photon collecting areas will make routine high precision spectropolarimetry feasible, 
such that even in the presence of thermal broadening, subtle changes in the 
polarization and PA may become measurable.

For cooler stars, dust may be the principal polarigenic agent.
Although the matrices for Mie and Rayleigh scattering (as used in our Monte Carlo modeling) 
are different, both favour forward and 
backward scattering, such that the differences between our predictions and those for Mie scattering are 
expected to be only qualitative. Furthermore, dust grains are not expected to have large thermal velocities that would 
result in significant line broadening.
Another potential opacity source for which smearing is not expected to be significant is that 
of neutral hydrogen.

Added to the high demand on sensitivity, another key aspect is that 
of spectral resolution. 
Currently, the resolution that can be achieved on 
common-user optical instruments is limited to 
$R$ $\la$ 10 000. Vink et al. (2005a) checked whether their predicted PA changes 
would be resolvable when they degraded their model spectra to $R$ $=$ 10 000. They 
found that the instrumental resolution would not wipe out the predicted $QU$ profiles.
The greater limitation is therefore sensitivity. This is because in the handling of  
spectropolarimetric data there is usually degeneracy between sensitivity and spectral resolution. 
Currently, one generally re-bins pixels across a spectral line, 
to gain the signal needed to achieve the required polarimetric accuracy. 

We conclude that while there is already plenty of evidence of single PA rotations in 
T Tauri and Herbig Ae stars that are entirely consistent with disrupted disks, 
the sensitivity is typically too poor to allow for quantitative comparisons 
between models and data. In any case, no instance of a double PA flip has yet been positively 
identified. However, this must be seen as absence of evidence rather than evidence of absence 
until appreciably greater sensitivity (with S/N $>>$ 1000) becomes routine.

Note that many of the PA rotation amplitudes we have predicted correspond to 
changes of only 0.05-0.1\% in Stokes $U$ throughout the spectral line. 
Upon occasion, this is measurable with 
today's instrumentation provided the integration times are long enough. 
This is well worth the effort, since line polarimetry
can uniquely obtain combined constraints on disk inclination and inner hole radius, 
as shown in Fig.\,\ref{incl}.

\section{Future applications}

The future of linear line polarimetry seems may become even brighter, as there are 
many exciting future applications to be considered. To name just a few:\\

\begin{enumerate}
\item Infrared spectropolarimetry (Oudmaijer et al. 2005)
\item Larger samples \& Surveys 
\item Monitoring data
\item Extra-galactic stars
\end{enumerate}

With respect to (i) the (near) infrared, it might become possible not only to detect the innermost 
regions of the accretion disks around the obscured most massive young massive (O-type) stars, but also 
the earliest phases of low- and intermediate mass star formation in pre-T Tauri and pre-Herbig systems.

With respect to the (ii) larger samples, it is noteworthy to realize that with large surveys, such as 
IPHAS (Drew et al. 2005) and VPHAS$+$ (Drew et al. 2014), thousands of young H$\alpha$ emitting stars are being discovered, allowing 
us to probe the mass-accretion physics as a function of mass, age, and environment (Barentsen et al. 2011; 2013; Kalari et al. in prep.).

\begin{figure}
\includegraphics[width=8cm]{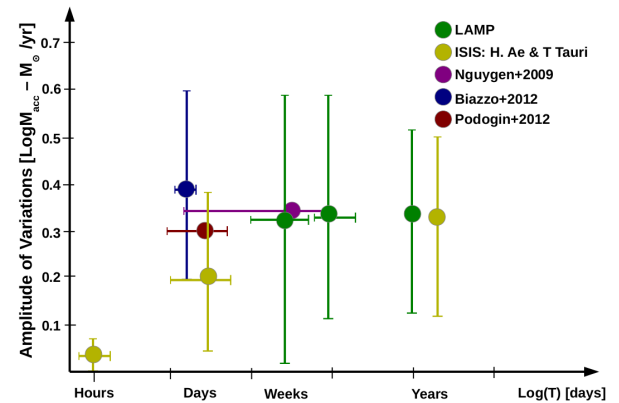}
\caption{
The amplitude in mass-accretion rate $\dot{M}_{\rm acc}$ variations 
for different timescales. Note that the maximum amplitude is reached on the 
(rotational) timescale (of days) and not on shorter timescales. On the other hand, 
sampling on significantly longer timescales does not seem to be required. See Costigan et al. (2014).
} 
\label{cost}
\end{figure}

\begin{figure*}
\includegraphics[width=16cm]{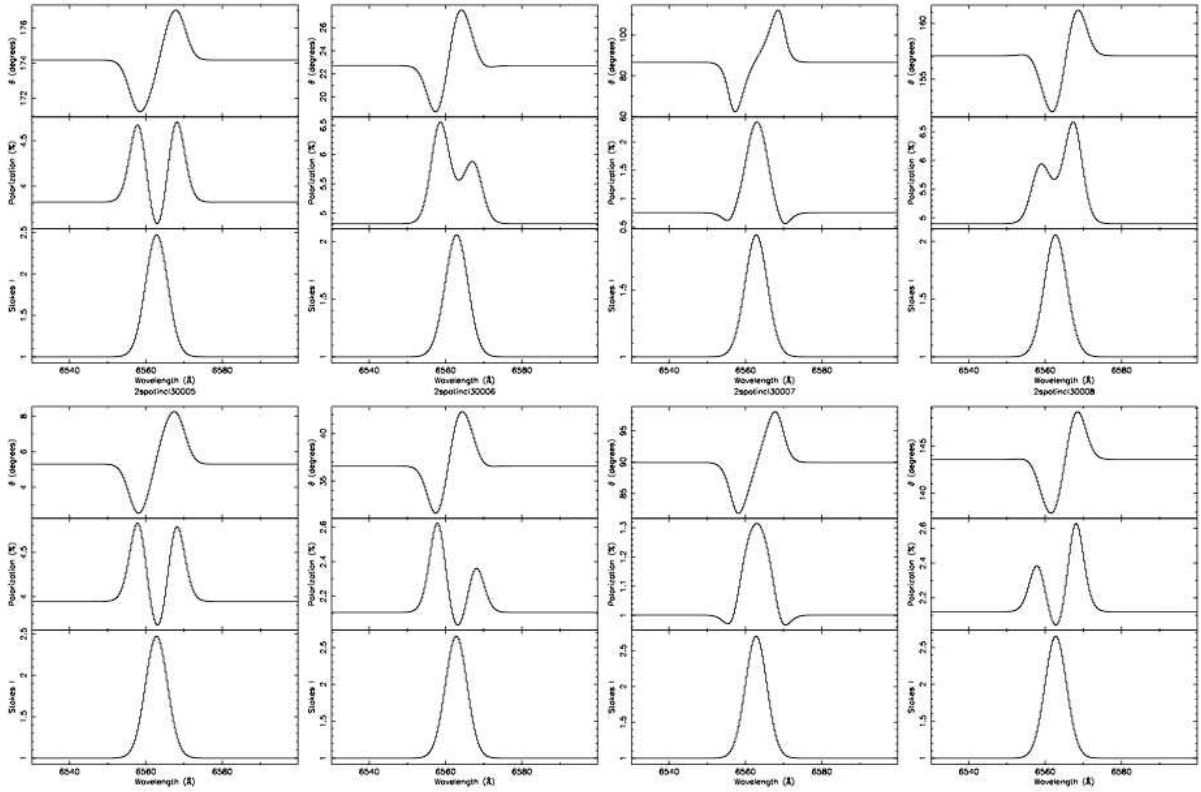}
\caption{
3D Monte Carlo simulation using {\sc torus} (Harries 2000) of a 
rotational modulation of 2 diametrically opposed spots on the stellar surface. 
The sequence consists of a full 360 degrees cycle. I.e. the eight snapshots from 0 to 315 degrees are 
taken every 45 degrees, i.e. at 0, 45, 90, 135, 180, 225, 270, and 315 degrees.
Note that the upper spot is initially (at 0 degrees) directed towards the observer.
}
\label{spots}
\end{figure*}

With respect to (iii) the monitoring, it has recently become clear that it is the rotational
timescale that is dominant in the mass-accretion physics in both T Tauri and Herbig Ae stars
(Costigan et al. 2014; see Fig.\,\ref{cost}). This means that linear spectropolarimetry modeling 
involving monitoring data such as predicted through Monte Carlo simulations 
shown in Fig.\,\ref{spots} can be employed to map the stellar-disk mass-accretion system in 3D.
This might become feasible with the space-based ultraviolet and 
optical spectropolarimeter Arago/UVMag\footnote{lesia.obspm.fr/UVMag} (PI Neiner).

Finally, with current 8m telescopes such as VLT, Herbig stars can already be discovered  in the 
low metallicity environment of the Large Magellanic Cloud (LMC). The mass-accretion rate ($\dot{M}_{\rm acc}$)
for the Herbig B[e] candidate VFTS\,822 (Kalari et al. 2014) seems particularly high, and in order 
to find out whether such rates are realistic we need the appropriate 3D data that will only be possible 
with linear spectropolarimetry. 

Currently, the main limitation is still sensitivity, but we are living in 
exciting times where the possibility of extremely large 40m telescopes (ELTs) is about to become reality. 
If the ELTs materialize with the badly needed polarization optics, we might 
be able to obtain spectropolarimetric data at a level of precision that has been feasible 
with 1D Stokes $I$ data for more than a century. 

It is really important to note that 3D Monte Carlo radiative transfer 
is well able to perform the required modeling, but the main limitation is still the lack of accurate 3D data! 

\section{Conclusions}

\begin{itemize}
\item Herbig Ae/Be stars have accretion disks on the smallest spatial scales.

\item Linear spectropolarimetry data are entirely consistent with disk accretion and fragmentation.

\item There is a {\it transition} in mass-accretion physics between Herbig Ae and Herbig Be stars.

\item The rotational timescale is key to changes in $\dot{M}_{\rm acc}$.

\item We {\it require} linear Stokes $QU$ monitoring to map the 3D geometry.

\end{itemize}

\acknowledgments
I would like to thank my long term specpol collaborators Rene Oudmaijer, Tim Harries, and Janet Drew, 
as well as my recent PMS PhD students Geert Barentsen, Grainne Costigan, and Venu Kalari for there 
input in mass-accretion topics in more recent times.

\nocite{*}
\bibliographystyle{spr-mp-nameyear-cnd}
\bibliography{biblio-u1}

\end{document}